# Everettian relative states in the Heisenberg picture

By Samuel Kuypers and David Deutsch

The Clarendon Laboratory, University of Oxford, Oxford OX1 3PU, UK

September 2020

## Abstract

Everett's relative-state construction in quantum theory has never been satisfactorily expressed in the Heisenberg picture. What one might have expected to be a straightforward process was impeded by conceptual and technical problems that we solve here. The result is a construction which, unlike Everett's one in the Schrödinger picture, makes manifest the locality of Everettian multiplicity, and its inherently approximative nature, and its origin in certain kinds of entanglement and locally inaccessible information. Our construction also allows us to give a more precise definition of an Everett 'universe', under which it is fully quantum, not quasi-classical, and we compare the Everettian decomposition of a quantum state with the foliation of a spacetime.

## Introduction

The dynamical evolution of all quantum systems is unitary at all times except – according to most traditional 'interpretations' of quantum theory – during measurements. The arbitrariness, non-locality, vagueness, inelegance and anthropocentricity of this supposed exception has caused many misconceptions and confusions both in theoretical physics and in philosophy, despite the work of Everett (1973), including his relative-state construction, which eliminated the exception. His primary technical innovation was to analyse quantum measurement processes by treating not only the system being measured but also the measuring apparatus or instrument (or *measurer,* for short) jointly as a unitarily evolving quantum system. As a consequence of that unitarity, when a measurer has measured an observable of another system, both the system and the measurer then exist in multiple instances such that every possible measurement outcome is observed in reality, but in autonomously evolving parts of reality, often called 'parallel universes'.[1]

Everett formulated his construction in the Schrödinger picture, in which the parallel universes are described by autonomously evolving components of the universal state vector. But the construction has never been satisfactorily expressed in the Heisenberg picture. This is potentially problematic for Everettian quantum theory. If, for some reason, relative states *could not* be formulated in the Heisenberg picture, that might indicate that Everettian quantum theory itself is not viable. Or if a successor theory to quantum theory, such as quantum gravity, were to lack a Schrödinger picture, it would not be possible to construct an Everettian interpretation for that theory, and all those traditional problems might reappear. For instance, Dirac (1965) cast doubt on the universal validity of the Schrödinger picture, arguing that there exist 'reasonable field theories in the Heisenberg picture … which do not allow of solutions of the wave equation to represent physical states in the Schrödinger picture'. Furthermore, the theory of relative states itself, whether by that name or not, is central to all versions of the quantum theory of measurement, for the relative states are where the outcomes of measurements appear explicitly.

## 1. Relative states in the Schrödinger picture

In the Schrödinger picture, dynamical evolution is described by the motion of the universal state vector in Hilbert space, and the quantum-mechanical observables are fixed Hermitian operators on that space. So in the Schrödinger picture relative-state construction, every possible outcome of the measurement of an observable appears, after a measurement, in a component of the universal state vector, and each such component consists of a product of two factors: an eigenstate of the measured observable; and a state of the measurer in which there is a record of the measured eigenvalue. When these components do not subsequently undergo quantum interference, the linearity of quantum dynamics ensures that each of them evolves independently, i.e. exactly as it would if the others were not there.

---

[1] Despite the name, an Everettian 'parallel universe' only ever has a finite spatial extent, because the Everettian multiplicity arises from, and spreads via, local interactions.

For example, consider a model quantum measurement in which a measurer, initially in a Schrödinger-picture state $|M\rangle \in \mathcal{H}_M$, measures the computational variable $\hat{q}_z$ of a qubit in the superposition $\frac{1}{\sqrt{2}}(|1\rangle + |-1\rangle) \in \mathcal{H}_S$ of the two eigenstates of $\hat{q}_z$. At the initial time $t$, the composite system $\mathcal{H}_M \otimes \mathcal{H}_S$ is in the state $|\Psi(t)\rangle = \frac{1}{\sqrt{2}}|M\rangle(|1\rangle + |-1\rangle)$, and the measurement, ending at time $t+1$ and resulting in the state $|\Psi(t+1)\rangle$, effects the transformation:

$$\frac{1}{\sqrt{2}}|M\rangle(|1\rangle + |-1\rangle) \xrightarrow{\text{measurement}} \frac{1}{\sqrt{2}}(|M_1\rangle|1\rangle + |M_{-1}\rangle|-1\rangle). \quad (1.1)$$

Here $|M_1\rangle$ and $|M_{-1}\rangle$ (not necessarily distinct from or orthogonal to $|M\rangle$) are orthonormal eigenstates of a suitable *pointer* observable $\hat{M}$ of the measurer, whose eigenvalues $M_1$ and $M_{-1}$ are assigned the meanings 'it was a 1' and 'it was a $-1$' respectively. The fact that $|M_1\rangle$ and $|M_{-1}\rangle$ are orthogonal, and that the final state has no components $|M_1\rangle|-1\rangle$ or $|M_{-1}\rangle|1\rangle$, is a token of the idealised accuracy of the measurement.

Provided there is no subsequent interference between $|M_1\rangle|1\rangle$ and $|M_{-1}\rangle|-1\rangle$ – for instance, if either the system or the measurer decoheres – the two instances of the combined system described by those two constituents will evolve independently after $t+1$ (cf. for instance Wallace (2011, 2012a); Zurek (1982)).

The term 'relative state' can be used in three ways: (i) $|M_1\rangle$ is the relative state of the measurer (relative to the eigenvalue 1 of the observable $\hat{q}_z$ of the qubit), and likewise for $|M_{-1}\rangle$; (ii) $|1\rangle$ is the relative state of the qubit relative to the eigenvalue $M_1$ of $\hat{M}$), and likewise for $|-1\rangle$; and (iii) the product $|M_1\rangle|1\rangle$ is the relative state of the combined system in which the outcome of the measurement was 1 (relative to the eigenvalues $M_1$ and 1 of the respective observables), and likewise for $M_{-1}$ and $-1$.

Since the measurer, initially in state $|M\rangle$, measures the computational variable $\hat{q}_z$, the relative states $|M_1\rangle|1\rangle$ and $|M_{-1}\rangle|-1\rangle$ of (1.1) are eigenstates of $\hat{q}_z$. These eigenstates have associated projectors

$$\hat{P}_1(\hat{q}_z) \stackrel{\text{def}}{=} \tfrac{1}{2}(\hat{1} + \hat{q}_z), \qquad \hat{P}_{-1}(\hat{q}_z) \stackrel{\text{def}}{=} \tfrac{1}{2}(\hat{1} - \hat{q}_z), \quad (1.2)$$

with $\hat{1}$ is the unit observable of $\mathcal{H}_M \otimes \mathcal{H}_S$. Each projector in (1.2) has an argument, namely the observable $\hat{q}_z$, and a subscript, which is an eigenvalue of $\hat{q}_z$ that specifies which eigenstate the projector projects onto.[2] Note that the sum of projectors (1.2) equals the unit observable $\hat{1}$, so (1.2) forms a so-called projection-valued measure (PVM).

The relative states are eigenstates of $\hat{q}_z$, so we express each relative state of the model measurement (1.1) as a (normalised) component of the state vector $|\Psi(t+1)\rangle$ in the direction specified by a projector of PVM (1.2), e.g. the relative state that contains the 'it was a 1' state $|M_1\rangle$ is

$$\frac{\hat{P}_1(\hat{q}_z)|\Psi(t+1)\rangle}{|\hat{P}_1(\hat{q}_z)|\Psi(t+1)\rangle|} = |M_1\rangle|1\rangle. \tag{1.3}$$

Where the denominator ensures that (1.3) is normalised. Since the elements of a PVM sum to unity, the state vector, after a quantum measurement, can invariably be expressed as a sum of relative states, each with a complex coefficient. It is a general feature of Everett's construction that the relative states after *any* perfect measurement of a computational variable can always be defined by a PVM in this manner.[3]

The projectors (1.2) have to act on the *full* state vector in $\mathcal{H}_M \otimes \mathcal{H}_S$ to define the relative states, and similarly, in situations where the multiverse consists not just of a measurer and a measured observable but of other systems, too, the multiverse is again decomposed into relative states through projectors acting on the full state vector. Thus, seemingly, the relative-state construction is non-local: a measurement of one system by another appears to affect the full multiverse instead of only the sub-systems that participated in the measurement. Indeed, some authors – e.g. Vaidman (2002) – have thereby been led to conclude that an Everett universe must be spatially unbounded. But this is merely due to a shortcoming of the Schrödinger picture. As we shall show in §3, the Heisenberg relative state construction is in every respect local and its 'universes' are always spatially bounded. Incidentally, this is in agreement with Everett's own understanding of his relative-state construction.[4]

---

[2] A projector for a degenerate eigenvalue projects onto an *eigenspace*, which is a vector space spanned by the eigenstates that correspond to a given (degenerate) eigenvalue. Measurements of degenerate eigenvalues are described by a straightforward generalisation of the model measurement (1.1).
[3] Note that imperfect measurements are described in a similar manner by a so-called positive operator valued measure, i.e. a set of semi-definite positive operators that sum to unity (Wallace (2012b)).

In this paper, we formulate Everett's construction within the Heisenberg picture, which is important because, for instance, information flow in quantum systems can only be analysed transparently in the Heisenberg picture (Deutsch & Hayden (2000), Deutsch (2011)). A construction similar to ours has been created by Horsman & Vedral (2007a), in their Section V. However, they do not draw the conclusion that the Heisenberg relative-states are states of fully quantum systems, with their own algebras of observables and with the identical structure to those of the original system; nor do they explain that these systems are always local in space and are therefore not literally universes.

## 2. Quantum computation in the Heisenberg picture

In contrast to the formalism of the Schrödinger picture, that of the Heisenberg picture consists of *evolving* q-number observables and a *fixed* state vector (or in the case of subsystems, a fixed density operator). Due to this difference, certain important properties of quantum systems are transparent in the Heisenberg picture but opaque in the Schrödinger picture. In particular, in the Heisenberg picture, the time-varying entities, the Heisenberg observables, are local, and affect each other only locally, just as in classical physics, unlike the time-varying state vector of the Schrödinger picture, which is non-local. This implies that in the Heisenberg picture, all physical information is assigned a location, even in entangled systems – see Deutsch & Hayden (*op. cit.*).

We shall construct the Heisenberg relative states for a quantum computational network; this technique, as used by Deutsch & Hayden (2000), ensures that our results about relative states generalise to all quantum systems. This is not because quantum computational networks are the most general system – a network consists exclusively of qubits – but since such networks are universal in their capacity to simulate with arbitrary accuracy any other physical system, any result obtained about relative states in these networks must be a general feature of quantum theory.

---

[4] Everett (1973) wrote that in his Many-Worlds, "*it is not so much the system which is affected by an observation as the observer, who becomes correlated with the system.* [Everett's italics. By 'correlated' he means entangled.] Only the observer state has changed, so as to become correlated with the state of the near system and hence naturally with the state of a remote system also."

Accordingly, consider a quantum computational network $\mathfrak{N}$ consisting of $n$ interacting qubits $\mathfrak{Q}_1, \mathfrak{Q}_2, \ldots, \mathfrak{Q}_n$. Following Gottesman (1999) and Deutsch & Hayden (*op. cit.*) we represent each qubit at the initial time $t$ as a triple of its observables, which we call the *descriptors* of the qubit:

$$\hat{\boldsymbol{q}}_a^{(t)} = \left( \hat{q}_{ax}^{(t)}, \; \hat{q}_{ay}^{(t)}, \; \hat{q}_{az}^{(t)} \right). \tag{2.1}$$

They satisfy the algebraic relations

$$\left. \begin{array}{ll} \hat{q}_{ai}^{(t)} \hat{q}_{aj}^{(t)} = \delta_{ij} \hat{1} + i\epsilon_{ij}{}^k \hat{q}_{ak}^{(t)} & (i, j, k \in \{x, y, z\}), \\ \left[ \hat{\boldsymbol{q}}_a^{(t)}, \hat{\boldsymbol{q}}_b^{(t)} \right] = 0 & (a \neq b). \end{array} \right\} \tag{2.2}$$

When an index appears twice in a product, once as a superscript and once as a subscript, as $k$ does in (2.2), we implicitly sum it over all its possible values, in accordance with the Einstein summation convention. Here, the values summed over are $\{x, y, z\}$.

From (2.2), it follows that for a network of $n$ qubits, each qubit descriptor can be expressed as a Hermitian $2^n \times 2^n$ matrix, so, for instance, the descriptors of a network consisting of a single qubit can be taken to have the initial matrix representation

$$\hat{q}_{1x}^{(0)} = \hat{\sigma}_x \stackrel{\text{def}}{=} \begin{pmatrix} 0 & 1 \\ 1 & 0 \end{pmatrix}; \quad \hat{q}_{1y}^{(0)} = \hat{\sigma}_y \stackrel{\text{def}}{=} \begin{pmatrix} 0 & -i \\ i & 0 \end{pmatrix}; \quad \hat{q}_{1z}^{(0)} = \hat{\sigma}_z \stackrel{\text{def}}{=} \begin{pmatrix} 1 & 0 \\ 0 & -1 \end{pmatrix}. \tag{2.3}$$

For the $n$-qubit network $\mathfrak{N}$, the initial representation (2.3) generalises to

$$\hat{\boldsymbol{q}}_a^{(0)} = \hat{1}^{a-1} \otimes \hat{\boldsymbol{\sigma}} \otimes \hat{1}^{n-a}, \tag{2.4}$$

where $\hat{1}^k$ is a tensor product of $k$ single-qubit identity operators, and $\hat{\boldsymbol{\sigma}} = (\hat{\sigma}_x, \hat{\sigma}_y, \hat{\sigma}_z)$. The initial matrix representation (2.4) of the descriptors will be altered by the subsequent evolution of the network, but since the descriptors of $\mathfrak{N}$ evolve unitarily, the algebra (2.2) is a constant of the motion.

It is convenient and conventional, and loses no generality, to assume that each gate in a network requires exactly 1 unit of time to act, so that the computational state of the network need only be specified at integer values of time. The effect of a general quantum gate **G** on $\mathfrak{Q}_a$ after one such period is

$$\widehat{\boldsymbol{q}}_a^{(t+1)} = U_\mathbf{G}^\dagger\left(\widehat{\boldsymbol{q}}_1^{(t)}, \dots, \widehat{\boldsymbol{q}}_n^{(t)}\right) \widehat{\boldsymbol{q}}_a^{(t)} U_\mathbf{G}\left(\widehat{\boldsymbol{q}}_1^{(t)}, \dots, \widehat{\boldsymbol{q}}_n^{(t)}\right). \tag{2.5}$$

Here the unitary $U_\mathbf{G}$ implements the gate **G** and is a function of the time-dependent descriptors of $\mathfrak{N}$. Note that in the Heisenberg picture, $U_\mathbf{G}$ generally varies with time and therefore its matrix representation in our basis changes with time as well, but it is a fixed (i.e. characteristic of the gate) function of the time-dependent descriptors. For example, the **not** gate is a single-qubit gate which, when it acts on qubit $\mathfrak{Q}_a$, toggles the value of its $z$-observable when it is sharp; the definition of the **not** gate in terms of the descriptors is

$$U_{\mathbf{not},a}\left(\widehat{\boldsymbol{q}}_1^{(t)}, \dots, \widehat{\boldsymbol{q}}_n^{(t)}\right) = \hat{q}_{ax}^{(t)}. \tag{2.6}$$

The only gate whose matrix representation in the Heisenberg picture is time-independent is the 'unit wire' **I**, the effect of which is that it leaves the descriptors of $\mathfrak{N}$ unchanged, and for a network of $n$ qubits, has matrix representation $U_\mathbf{I} = \hat{1}^n$.

Since the unitary of a gate can be expressed as a characteristic *function* of the descriptors, a gate also has a characteristic *effect* on those descriptors, which can be specified algebraically. This provides a complementary way of defining a gate in the Heisenberg picture. For example, using (2.5) and (2.6), we find that after the application of the **not** gate, the descriptors of $\mathfrak{Q}_a$ at time $t+1$ have the following fixed expression in terms of its descriptors at time $t$:

$$\mathbf{not}: \quad \widehat{\boldsymbol{q}}_a^{(t+1)} = \left(\hat{q}_{ax}^{(t)}, -\hat{q}_{ay}^{(t)}, -\hat{q}_{az}^{(t)}\right), \tag{2.7}$$

while all other qubits in the network are unaffected. Expressing gates in form (2.7) rather than (2.6) often makes their effect on a network more transparent. For example, it is manifest in (2.7) that the eigenvalues of the z-observable of $\mathfrak{Q}_a$ are toggled by the **not** gate, as $\hat{q}_{az}^{(t+1)} = -\hat{q}_{az}^{(t)}$.

$\mathfrak{N}$ is fully specified by its time-dependent descriptors and constant Heisenberg state $|\Psi\rangle$. We shall fix the Heisenberg state to be a standard constant by using the basis $\{|\hat{q}_{1z}, \ldots, \hat{q}_{nz}; t\rangle\}$ of instantaneous eigenstates of the descriptors $\{\hat{q}_{az}^{(t)}\}$ so that $|\Psi\rangle$ equals $|1, \ldots, 1; t = 0\rangle$. Then the expectation value of an arbitrary time-dependent observable $\hat{A}(t)$ of $\mathfrak{N}$ is

$$\langle \hat{A}(t) \rangle \overset{\text{def}}{=} \langle \Psi | \hat{A}(t) | \Psi \rangle \overset{\text{def}}{=} \langle 1, \ldots, 1; 0 | \hat{A}(t) | 1, \ldots, 1; 0 \rangle. \tag{2.8}$$

Since the Heisenberg 'state' $|\Psi\rangle$ is a fixed constant, it contains no information about the physical state of $\mathfrak{N}$. All such information resides instead in the Heisenberg observables of $\mathfrak{N}$.

Whenever the expectation value of an observable $\hat{A}(t)$ of $\mathfrak{N}$ has the property that $\langle \hat{A}(t) \rangle^2 = \langle \hat{A}(t)^2 \rangle$, the Heisenberg state is an eigenstate of $\hat{A}(t)$ and the observable is then said to be *sharp* at time $t$ with the value $\langle \hat{A}(t) \rangle$ (in Everettian terms, it has 'the same value in all universes' at time $t$), where that value is one of the eigenvalues of $\hat{A}(t)$.

In the Heisenberg picture, two qubits $\mathfrak{Q}_a$ and $\mathfrak{Q}_b$ are entangled at time $t$ if there exists a pair of descriptors $\hat{q}_{ai}^{(t)}$ and $\hat{q}_{bj}^{(t)}$ such that $\langle \hat{q}_{ai}^{(t)} \hat{q}_{bj}^{(t)} \rangle \neq \langle \hat{q}_{ai}^{(t)} \rangle \langle \hat{q}_{bj}^{(t)} \rangle$ (Horsman & Vedral (2007b)). An entangled quantum system is capable of storing *locally inaccessible information* – information that cannot be recovered through any measurement of that quantum system alone (Deutsch & Hayden *op. cit.*). This fact is transparent in the Heisenberg picture, since it makes the location of all physical information explicit. For instance, consider two entangled qubits $\mathfrak{Q}_a$ and $\mathfrak{Q}_b$ that, at the initial time $t$, have expectation values $\langle \hat{q}_a^{(t)} \rangle = (0,0,0)$ and $\langle \hat{q}_b^{(t)} \rangle = (0,0,0)$ and $\langle \hat{q}_{az}^{(t)} \hat{q}_{bz}^{(t)} \rangle = 1$, so the qubits are in a Bell state. Then, at time $t$, $\mathfrak{Q}_a$ effects a gate $R_z(\theta)$, which rotates $\mathfrak{Q}_a$ by an angle of $\theta$ radians about its z-axis. That is to say, the descriptors of $\mathfrak{Q}_a$ evolve as follows:

$$\hat{\boldsymbol{q}}_a^{(t+1)} = \left( \cos\theta \, \hat{q}_{ax}^{(t)} - \sin\theta \, \hat{q}_{ay}^{(t)}, \; \cos\theta \, \hat{q}_{ay}^{(t)} + \sin\theta \, \hat{q}_{ax}^{(t)}, \; \hat{q}_{az}^{(t)} \right), \tag{2.9}$$

Manifestly, the information about the angle $\theta$ is located in the $x$- and $y$-observables of $\hat{\boldsymbol{q}}_a^{(t+1)}$, yet none of the expectation values of the qubit's observables reflect the fact that it contains this information since

$$\left\langle \hat{\boldsymbol{q}}_a^{(t+1)} \right\rangle = (0,0,0). \tag{2.10}$$

Thus, the angle $\theta$ is *locally inaccessible* – i.e. the results of measurements made on $\mathfrak{Q}_a$ alone, by measurers with no prior knowledge of $\theta$, reveal nothing about $\theta$, even statistically (if the whole experiment preparing $\mathfrak{Q}_a$ is repeated any number of times). As we shall show in §4, thanks to entanglement, instances of a system in a relative state are also capable of storing locally inaccessible information.

### 3. The Heisenberg picture relative-state construction

When Everett's construction is formulated in the Schrödinger picture, the joint state vector of the observed system and the measurer are decomposed into *relative state vectors*, each describing a 'universe'.[5] Though translating between the Schrödinger and Heisenberg pictures is usually a straightforward process, doing so for the relative-state construction brings with it the conceptual difficulty of decomposing the Heisenberg descriptors into *relative descriptors*: q-numbers which, following a measurement, represent individual instances of a combined quantum system, and which should correspond to the Schrödinger-picture relative-state vectors. Our construction in what follows resolves that difficulty.

Consider a measurement performed by a network $\mathfrak{M}$ that consists of the two qubits $\mathfrak{Q}_M$ and $\mathfrak{Q}_S$, where $\mathfrak{Q}_M$ represents an idealised measurer that is programmed to measure the $z$-observable of system $\mathfrak{Q}_S$. The qubits in $\mathfrak{M}$ have initial ($t = 0$) matrix representation (2.4) with $n = 2$:

---

[5] We call these 'universes' to match the terminology used by DeWitt & Graham (1973) and Everett (see Byrne (2010)). Other authors have used different terms to refer to these Everettian entities, such as 'worlds', 'branches' or 'histories'. The different metaphysical connotations of those terms do not concern use here: our construction would apply to any of them. And the Heisenberg picture makes explicit that these 'universes', 'worlds', 'branches' or 'histories' are local: as we shall show, immediately after a measurement the entities that exist in multiple relative states are precisely the system and the measuring instrument, and nothing beyond.

$$\left.\begin{aligned}\hat{\boldsymbol{q}}_S^{(0)} &= (\hat{\sigma}_x \otimes \hat{1},\ \hat{\sigma}_y \otimes \hat{1},\ \hat{\sigma}_z \otimes \hat{1}), \\ \hat{\boldsymbol{q}}_M^{(0)} &= (\hat{1} \otimes \hat{\sigma}_x,\ \hat{1} \otimes \hat{\sigma}_y,\ \hat{1} \otimes \hat{\sigma}_z),\end{aligned}\right\} \tag{3.1}$$

To reduce notational clutter, we shall abbreviate this as

$$\left.\begin{aligned}\hat{\boldsymbol{q}}_S &= (\hat{q}_{Sx},\ \hat{q}_{Sy},\ \hat{q}_{Sz}), \\ \hat{\boldsymbol{q}}_M &= (\hat{q}_{Mx},\ \hat{q}_{My},\ \hat{q}_{Mz}).\end{aligned}\right\} \tag{3.2}$$

The Heisenberg state $|\Psi\rangle$ of the network equals $|1,1;0\rangle$, so that at $t = 0$, the $z$-observables of both $\mathfrak{Q}_M$ and $\mathfrak{Q}_S$ are sharp with value 1. Everett's relative-state construction explains why a unitarily evolving measurer appears to observe only one of the eigenvalues of an observable that is non-sharp: the other eigenvalues occurred in different 'universes'. To elaborate this, we apply a Hadamard gate **H** to $\mathfrak{Q}_S$ before the measurement begins. **H** is a single-qubit gate which rotates a qubit by $\pi$ radians about the axis that bisects the angle between the $x$- and $z$-axis. That is to say, when **H** is applied to $\mathfrak{Q}_S$, it has the effect

$$\left.\begin{aligned}\hat{\boldsymbol{q}}_S^{(1)} &= (\hat{q}_{Sz},\ -\hat{q}_{Sy},\ \hat{q}_{Sx}), \\ \hat{\boldsymbol{q}}_M^{(1)} &= (\hat{q}_{Mx},\ \hat{q}_{My},\ \hat{q}_{Mz}).\end{aligned}\right\} \tag{3.3}$$

Thus at $t = 1$, the Heisenberg state is not an eigenstate of $\hat{q}_{Sz}^{(1)}$, so this descriptor has no sharp value.

Next, between $t = 1$ and $t = 2$, $\mathfrak{Q}_M$ performs an idealised measurement of the $z$-observable of $\mathfrak{Q}_S$. This is implemented by the two-qubit 'controlled-not' (or 'perfect measurement') gate **cnot**. The qubits that this gate acts on are known as the *control* and *target* qubits, which in this case are $\mathfrak{Q}_S$ and $\mathfrak{Q}_M$ respectively. The effect of a **cnot** gate depends on the value of the control qubit: at a general time $t$, **cnot** has no effect on the target if $\hat{q}_{Sz}^{(t)}$ of the control is sharp with value 1 and performs a **not** operation on the target if $\hat{q}_{Sz}^{(t)}$ of the control is sharp with value $-1$. It follows that the unitary $U_{\mathbf{cnot}}$ that implements the **cnot** operation at a general time $t$ is

$$U_{\mathbf{cnot}}\left(\hat{\boldsymbol{q}}_S^{(t)}, \hat{\boldsymbol{q}}_M^{(t)}\right) = \hat{P}_1\left(\hat{q}_{Sz}^{(t)}\right) + U_{\mathbf{not},M}\left(\hat{\boldsymbol{q}}_M^{(t)}\right)\hat{P}_{-1}\left(\hat{q}_{Sz}^{(t)}\right). \tag{3.4}$$

The operators $\hat{P}_1\left(\hat{q}_{Sz}^{(t)}\right) = \frac{1}{2}\left(\hat{1} + \hat{q}_{Sz}^{(t)}\right)$ and $\hat{P}_{-1}\left(\hat{q}_{Sz}^{(t)}\right) = \frac{1}{2}\left(\hat{1} - \hat{q}_{Sz}^{(t)}\right)$ are the Heisenberg equivalent of the projectors of (1.2), and their appearance in (3.4) makes explicit that **cnot** is a conditional **not** operation.

By using (2.5) and (3.4), we find that the effect of the **cnot** gate on the descriptors of the two qubits at a general time $t$ is

$$\left\{\begin{matrix}\hat{\mathbf{q}}_S^{(t+1)} \\ \hat{\mathbf{q}}_M^{(t+1)}\end{matrix}\right\} = \left\{\begin{matrix}\left(\hat{q}_{Sx}^{(t)}\hat{q}_{Mx}^{(t)}, & \hat{q}_{Sy}^{(t)}\hat{q}_{Mx}^{(t)}, & \hat{q}_{Sz}^{(t)}\right) \\ \left(\hat{q}_{Mx}^{(t)}, & \hat{q}_{My}^{(t)}\hat{q}_{Sz}^{(t)}, & \hat{q}_{Mz}^{(t)}\hat{q}_{Sz}^{(t)}\right)\end{matrix}\right\}. \tag{3.5}$$

If both descriptors $\hat{q}_{Sz}^{(t)}$ and $\hat{q}_{Mz}^{(t)}$ of (3.5) are initially sharp and, in particular, if $\left\langle\hat{q}_{Mz}^{(t)}\right\rangle = 1$, then the **cnot** evolves the value of $\hat{q}_{Mz}^{(t)}$ into a 'copy' of the value of $\hat{q}_{Sz}^{(t)}$ – that is to say $\left\langle\hat{q}_{Mz}^{(t+1)}\right\rangle = \left\langle\hat{q}_{Sz}^{(t)}\right\rangle = \pm 1$. Accordingly, after the application of **cnot**, the sharp values 1 and $-1$ of $\hat{q}_{Mz}^{(t)}$ have the unequivocal physical meaning 'the control was a 1' and 'the control was a $-1$.' If $\hat{q}_{Sz}^{(t)}$ is non-sharp, and $\hat{q}_{Mz}^{(t)}$ is sharp with value 1, then a **cnot** operation still 'copies' the value of a superposed control onto $\hat{q}_{Mz}^{(t)}$ by evolving the target into a mixed state, and the combined system is then entangled.

Note in passing, by inspection of (3.5), that the **cnot** gate *also* effects a perfect measurement of the $x$-observable of $\mathfrak{Q}_M$, storing the result in the $x$-observable of $\mathfrak{Q}_S$, and in that interpretation of what the gate does, the roles of 'target' and 'control' are interchanged. In general, whether a system is being measured or performs a measurement depends not only on the interaction between them but also on how the information in each system is *subsequently* used, and in general, it cannot be freely chosen which information stored in the pair of qubits to use. Decoherence would, for example, determine which of the qubits in (3.5) can act as a measurer (Zurek (1981), (1982)).

Using (3.3) and (3.5), we find that the descriptors at $t = 2$, after the application of the **cnot** gate, are

$$\left.\begin{matrix}\hat{\mathbf{q}}_S^{(2)} = \left(\hat{q}_{Sz}\hat{q}_{Mx}, & -\hat{q}_{Sy}\hat{q}_{Mx}, & \hat{q}_{Sx}\right), \\ \hat{\mathbf{q}}_M^{(2)} = \left(\hat{q}_{Mx}, & \hat{q}_{My}\hat{q}_{Sx}, & \hat{q}_{Mz}\hat{q}_{Sx}\right).\end{matrix}\right\} \tag{3.6}$$

The Schrödinger state vector of the qubits at $t = 2$ is

$$|\Psi(2)\rangle = \frac{1}{\sqrt{2}}(|1\rangle|1\rangle + |-1\rangle|-1\rangle), \tag{3.7}$$

with relative states $|1\rangle|1\rangle$ and $|-1\rangle|-1\rangle$. To construct the corresponding relative state decomposition *of the descriptors* (3.6), we use (3.4) to express the descriptors of $\mathfrak{Q}_M$ at $t = 2$ in terms of descriptors at $t = 1$:

$$\begin{aligned}\hat{\boldsymbol{q}}_M^{(2)} &= U_{\mathbf{cnot}}^\dagger\left(\hat{\boldsymbol{q}}_S^{(1)}, \hat{\boldsymbol{q}}_M^{(1)}\right)\hat{\boldsymbol{q}}_M^{(1)} U_{\mathbf{cnot}}\left(\hat{\boldsymbol{q}}_S^{(1)}, \hat{\boldsymbol{q}}_M^{(1)}\right) = \\ &\quad \hat{\boldsymbol{q}}_M^{(1)}\,\hat{P}_1\left(\hat{q}_{Sz}^{(1)}\right) + \left[U_{\mathbf{not},M}^\dagger\left(\hat{\boldsymbol{q}}_M^{(1)}\right)\hat{\boldsymbol{q}}_M^{(1)} U_{\mathbf{not},M}\left(\hat{\boldsymbol{q}}_M^{(1)}\right)\right]\hat{P}_{-1}\left(\hat{q}_{Sz}^{(1)}\right).\end{aligned} \tag{3.8}$$

(3.8) shows the **cnot** 'splitting' $\mathfrak{Q}_M$ into two simultaneously existing *instances*: one of them participates in a **not** operation, while the second is left unchanged. Let us call these two instances $\mathfrak{Q}_{M,1}$ and $\mathfrak{Q}_{M,-1}$, where the additional subscript specifies whether the instance is relative to the measured value 1 or $-1$. The instances $\mathfrak{Q}_{M,1}$ and $\mathfrak{Q}_{M,-1}$ are in the respective Heisenberg-picture relative states and are defined as

$$\left.\begin{aligned}\mathfrak{Q}_{M,1}: \quad & \hat{\boldsymbol{q}}_{M,1}^{(2)} \stackrel{\text{def}}{=} \hat{\boldsymbol{q}}_M^{(2)} \hat{P}_1\left(\hat{q}_{Sz}^{(2)}\right), \\ \mathfrak{Q}_{M,-1}: \quad & \hat{\boldsymbol{q}}_{M,-1}^{(2)} \stackrel{\text{def}}{=} \hat{\boldsymbol{q}}_M^{(2)} \hat{P}_{-1}\left(\hat{q}_{Sz}^{(2)}\right).\end{aligned}\right\} \tag{3.9}$$

Because of relations (2.2), the projectors $\hat{P}_1\left(\hat{q}_{Sz}^{(2)}\right)$ and $\hat{P}_{-1}\left(\hat{q}_{Sz}^{(2)}\right)$ commute with $\hat{\boldsymbol{q}}_M^{(2)}$. This ensures that the order of the factors in (3.9) makes no difference and that $\hat{\boldsymbol{q}}_{M,1}^{(2)}$ and $\hat{\boldsymbol{q}}_{M,-1}^{(2)}$ are Hermitian. Also, $\hat{q}_{Sz}^{(1)} = \hat{q}_{Sz}^{(2)}$, and so there is no discrepancy between the definitions of $\hat{\boldsymbol{q}}_{M,1}^{(2)}$ and $\hat{\boldsymbol{q}}_{M,-1}^{(2)}$ in (3.8) and (3.9).

$\mathfrak{Q}_{M,1}$ and $\mathfrak{Q}_{M,-1}$ each hold a record of a different eigenvalue of $\hat{q}_{Sz}$, which is why the instances $\mathfrak{Q}_{M,1}$ and $\mathfrak{Q}_{M,-1}$ are in distinct *histories* of the measurer, just as in the Schrödinger-picture relative state. Hence, we shall call the descriptors of $\mathfrak{Q}_{M,1}$ and $\mathfrak{Q}_{M,-1}$ the *relative descriptors* (or descriptors relative to the results 1 and $-1$, etc.).

The relative descriptors satisfy an appropriate form of the Pauli algebra (2.3). For example, note that the squares of the relative descriptors of $\mathfrak{Q}_{M,1}$ do not equal the unit observable $\hat{1}$ of the full algebra of operators since $\left(\hat{q}_{(M,1)i}\right)^2 = \hat{P}_1(\hat{q}_{Sz})$, and moreover, $\hat{P}_1(\hat{q}_{Sz})\,\hat{q}_{(M,1)i} = \hat{q}_{(M,1)i}\hat{P}_1(\hat{q}_{Sz}) = \hat{q}_{(M,1)i}$ for all $i \in \{x,y,z\}$. (Here we have dropped the time labels for clarity.) Because of these algebraic relations, $\hat{1}_{M,1} \stackrel{\text{def}}{=} \hat{P}_1(\hat{q}_{Sz})$ is the unit observable in the relative algebra of $\mathfrak{Q}_{M,1}$, and similarly, $\hat{1}_{M,-1} \stackrel{\text{def}}{=} \hat{P}_{-1}(\hat{q}_{Sz})$ is the unit observable in the relative algebra of $\mathfrak{Q}_{M,-1}$. Physically, this means that relative descriptors adhere to the Pauli algebra, and $\mathfrak{Q}_{M,1}$ and $\mathfrak{Q}_{M,-1}$ are, thus, fully-fledged qubits:

$$\left.\begin{array}{l}\hat{q}_{(M,1)i}\hat{q}_{(M,1)j} = \delta_{ij}\hat{1}_{M,1} + i\epsilon_{ij}{}^k\hat{q}_{(M,1)k},\\ \hat{q}_{(M,-1)i}\hat{q}_{(M,-1)j} = \delta_{ij}\hat{1}_{M,-1} + i\epsilon_{ij}{}^k\hat{q}_{(M,-1)k},\end{array}\right\} \qquad (i,j,k \in \{x,y,z\}). \tag{3.10}$$

So, in each Everett 'universe', quantum theory holds in full, but the matrix representation required to represent separately the sets of relative descriptors is of a smaller dimension than that of the full descriptors of $\mathfrak{Q}_M$.

Because $\mathfrak{Q}_S$ also contains a copy of $\hat{q}_{Mz}^{(2)}$, there is a similar relative state description of $\mathfrak{Q}_S$ relative to the values $1$ and $-1$ of $\hat{q}_{Mz}^{(2)}$. These instances of $\mathfrak{Q}_S$ are

$$\left.\begin{array}{ll}\mathfrak{Q}_{S,1}: & \hat{\boldsymbol{q}}_{S,1}^{(2)} \stackrel{\text{def}}{=} \hat{\boldsymbol{q}}_S^{(2)}\hat{P}_1\left(\hat{q}_{Mz}^{(2)}\right),\\ \mathfrak{Q}_{S,-1}: & \hat{\boldsymbol{q}}_{S,-1}^{(2)} \stackrel{\text{def}}{=} \hat{\boldsymbol{q}}_S^{(2)}\hat{P}_{-1}\left(\hat{q}_{Mz}^{(2)}\right).\end{array}\right\} \tag{3.11}$$

The order of the factors in (3.11) is again irrelevant since the observables of $\mathfrak{Q}_M$ and $\mathfrak{Q}_S$ commute. And the instances $\mathfrak{Q}_{S,1}$ and $\mathfrak{Q}_{S,-1}$ again obey the appropriate form of the Pauli algebra, but now with relative unit observables $\hat{1}_{S,1} \stackrel{\text{def}}{=} \hat{P}_1\left(\hat{q}_{Mz}^{(2)}\right)$ and $\hat{1}_{S,-1} \stackrel{\text{def}}{=} \hat{P}_{-1}\left(\hat{q}_{Mz}^{(2)}\right)$.

Analogously to the Schrödinger picture relative-state construction, the projectors $\hat{P}_1\left(\hat{q}_{Sz}^{(2)}\right)$ and $\hat{P}_{-1}\left(\hat{q}_{Sz}^{(2)}\right)$ of the measured observable form a PVM, and the relative descriptors (3.9) of $\mathfrak{Q}_M$ are defined as products of $\hat{\boldsymbol{q}}_M^{(2)}$ and the projectors of this PVM. Since quantum computational networks are universal, it must be a general feature of the Heisenberg relative-state construction that the relative descriptors of *any* measurer, after a perfect measurement, are the absolute descriptors projected via a PVM in the above manner. This resolves the conceptual difficulty of expressing the Heisenberg descriptors in terms of relative descriptors, and importantly, our solution is also in agreement with Horsman & Vedral (2007a).

Borrowing a term from differential geometry, we say that entangled systems, such as $\mathfrak{Q}_M$ and $\mathfrak{Q}_S$, can be *foliated* into relative states. In the Heisenberg picture, such a foliation is always spatially *local* since only those systems that are entangled due to a measurement are foliated into relative states, whereas no other systems are affected. This is evident from the fact that the foliations of $\mathfrak{Q}_M$ and $\mathfrak{Q}_S$ are individually specified, the latter in (3.11) and the former in (3.9). In contrast, the Schrödinger picture, which is not a local description of quantum systems, does not allow systems to be foliated individually: a composite system can only be foliated jointly by a projector operating on the full Schrödinger state vector. So, in the Heisenberg picture, unentangled systems cannot be foliated into physically meaningful relative states.

Foliating a quantum system changes the expectation value in relative states from those in the absolute network $\mathfrak{M}$. For example, at time $t = 2$, the conditional expectation value for an arbitrary descriptor $\hat{q}_{Mi}^{(2)}$ of $\mathfrak{Q}_M$ given that the *z*-observable of $\mathfrak{Q}_S$ was measured to be a 1 is

$$\left\langle \hat{q}_{Mi}^{(2)} \right\rangle_{M,1} \stackrel{\text{def}}{=} \frac{\left\langle \hat{q}_{Mi}^{(2)} \hat{P}_1 \left( \hat{q}_{Sz}^{(2)} \right) \right\rangle}{\left\langle \hat{P}_1 \left( \hat{q}_{Sz}^{(2)} \right) \right\rangle}. \tag{3.12}$$

Thus, so long as no further entangling or un-entangling interactions take place, the expectation values of $\mathfrak{Q}_{M,1}$ (and any other system in a relative state, relative to the result 1 of our measurement) are the same as they would be *if* the Heisenberg state had changed during the measurement:

$$|\Psi\rangle \to |\Psi_{M,1}\rangle \stackrel{\text{def}}{=} \frac{\hat{P}_1 \left( \hat{q}_{Sz}^{(2)} \right) |\Psi\rangle}{|\hat{P}_1 \left( \hat{q}_{Sz}^{(2)} \right) |\Psi\rangle|}. \tag{3.13}$$

We call $|\Psi_{M,1}\rangle$ the *relative Heisenberg state*, relative to the result 1 of the measurement of $\hat{q}_{Sz}^{(2)}$. Note that the normalisation factor $\left\langle \hat{P}_1 \left( \hat{q}_{Sz}^{(2)} \right) \right\rangle$ is necessarily non-zero, as $\hat{q}_{Sz}^{(2)}$ is non-sharp.

Since the expectation values of the relative descriptors differ from those of the observables of the absolute (multiversal) network $\mathfrak{M}$, it is possible for an observable, such as $\hat{q}_{Mz}^{(2)}$, not to be sharp with respect to the absolute network (i.e. $\langle \hat{q}_{Mz}^{(2)} \rangle = 0$) but for $\hat{q}_{(M,1)z}^{(2)}$ to be sharp *relative* to $\hat{q}_{Sz}^{(2)}$ having been measured with value 1 (i.e. $\langle \hat{q}_{(M,1)z}^{(2)} \rangle_{M,1} = 1$). We can describe this informally as $\hat{q}_{Mz}^{(2)}$ having a sharp value in each universe (or in each relative state) but not in the multiverse at large.

## 4. Everett 'universes'

The relative-state construction, in any picture, is a calculus for expressing a quantum system in terms of the behaviour of its constituent instances in situations where some or all of those instances evolve autonomously. In such situations the system consists of parallel 'histories' or 'universes', each represented by a relative state. However, in generic situations a foliation into such universes does not exist. Whether it does depends on the dynamical evolution of the entangled system. To study the types of dynamical evolution that permit an entangled system to be foliated into parallel 'universes', let us once more consider the quantum computational network $\mathfrak{M}$ at $t = 2$, when the network is foliated into $\mathfrak{Q}_{S,1}$ with $\mathfrak{Q}_{M,1}$, and $\mathfrak{Q}_{S,-1}$ with $\mathfrak{Q}_{M,-1}$. The gates that cause these relative states of $\mathfrak{M}$ to evolve autonomously will, thanks to the universality of quantum computational networks, also model when a general quantum system can be foliated into 'universes'.

$\mathfrak{Q}_M$ can be foliated into relative states (3.9) for the duration that it contains a copy of $\hat{q}_{Sz}^{(2)}$. $\mathfrak{Q}_M$ contains a copy of $\hat{q}_{Sz}^{(2)}$ at a general time $t$ if the product of one of $\mathfrak{Q}_M$'s components and $\hat{q}_{Sz}^{(2)}$ is sharp – for instance, the product $\hat{q}_{Mz}^{(2)} \hat{q}_{Sz}^{(2)}$ is sharp with value 1 – and $\mathfrak{Q}_M$ can then be decomposed in terms of $\mathfrak{Q}_{M,1}$ and $\mathfrak{Q}_{M,-1}$:

$$\left.\begin{aligned} \mathfrak{Q}_{M,1}: \quad & \hat{\boldsymbol{q}}_{M,1}^{(t)} \stackrel{\text{def}}{=} \hat{\boldsymbol{q}}_M^{(t)} \hat{P}_1\left(\hat{q}_{Sz}^{(2)}\right), \\ \mathfrak{Q}_{M,-1}: \quad & \hat{\boldsymbol{q}}_{M,-1}^{(t)} \stackrel{\text{def}}{=} \hat{\boldsymbol{q}}_M^{(t)} \hat{P}_{-1}\left(\hat{q}_{Sz}^{(2)}\right). \end{aligned}\right\} \quad (4.1)$$

Here $t \geq 2$. Similarly, qubit $\mathfrak{Q}_S$ can then be foliated into the relative states

$$\left.\begin{aligned} \mathfrak{Q}_{S,1}: \quad & \hat{\boldsymbol{q}}_{S,1}^{(t)} \stackrel{\text{def}}{=} \hat{\boldsymbol{q}}_S^{(t)} \hat{P}_1\left(\hat{q}_{Mz}^{(2)}\right), \\ \mathfrak{Q}_{S,-1}: \quad & \hat{\boldsymbol{q}}_{S,-1}^{(t)} \stackrel{\text{def}}{=} \hat{\boldsymbol{q}}_S^{(t)} \hat{P}_{-1}\left(\hat{q}_{Mz}^{(2)}\right). \end{aligned}\right\} \quad (4.2)$$

A qubit such as $\mathfrak{Q}_a$ is said to evolve autonomously between $t$ and $t+1$ if its descriptors $\hat{\boldsymbol{q}}_a^{(t+1)}$ depend only on $\hat{\boldsymbol{q}}_a^{(t)}$. Using this definition, which class of gates, acting between $t=2$ and $t=3$, would ensure that the relative states of $\mathfrak{M}$ evolve autonomously between those times? Evidently, not all gates do since, for instance, a second application of **cnot** would erase the information about $\mathfrak{Q}_S$ contained in $\mathfrak{Q}_M$, so this gate would destroy the relative-state structure. On the other hand, any gate that does not affect $\hat{q}_{Sz}^{(2)}$ and $\hat{q}_{Mz}^{(2)}$ will ensure that the relative states evolve autonomously as it will leave the information about $\hat{q}_{Sz}^{(2)}$ contained in $\hat{q}_{Mz}^{(2)}$ unaffected. As such, consider a gate **F**, which is implemented by the unitary $U_\mathbf{F}$, and which commutes with $\hat{q}_{Sz}^{(2)}$ and $\hat{q}_{Mz}^{(2)}$ so as to leave these descriptors unaltered. The most general expression of $U_\mathbf{F}$ is

$$U_\mathbf{F}\left(\hat{\boldsymbol{q}}_S^{(2)}, \hat{\boldsymbol{q}}_M^{(2)}\right) = \alpha \hat{1} + \beta \hat{q}_{Mz}^{(2)} + \gamma \hat{q}_{Sz}^{(2)} + \delta \hat{q}_{Mz}^{(2)} \hat{q}_{Sz}^{(2)}. \tag{4.3}$$

The coefficients $\alpha$, $\beta$, $\gamma$, and $\delta$ are real numbers, with the constraint that $U_\mathbf{F}$ is unitary. Using (4.3) and (4.2), the effect of gate **F** on the relative descriptors of $\mathfrak{Q}_{S,1}$ and $\mathfrak{Q}_{S,-1}$ is

$$\left.\begin{aligned}\hat{\boldsymbol{q}}_{S,1}^{(3)} &\stackrel{\text{def}}{=} \left(U_\mathbf{F}^\dagger\left(\hat{\boldsymbol{q}}_S^{(2)}, \hat{\boldsymbol{q}}_M^{(2)}\right) \hat{\boldsymbol{q}}_S^{(2)} U_\mathbf{F}\left(\hat{\boldsymbol{q}}_S^{(2)}, \hat{\boldsymbol{q}}_M^{(2)}\right)\right) \hat{P}_1\left(\hat{q}_{Mz}^{(2)}\right) = U_{\mathbf{F}'}^\dagger\left(\hat{\boldsymbol{q}}_{S,1}^{(2)}\right) \hat{\boldsymbol{q}}_{S,1}^{(2)} U_{\mathbf{F}'}\left(\hat{\boldsymbol{q}}_{S,1}^{(2)}\right), \\ \hat{\boldsymbol{q}}_{S,-1}^{(3)} &\stackrel{\text{def}}{=} \left(U_\mathbf{F}^\dagger\left(\hat{\boldsymbol{q}}_S^{(2)}, \hat{\boldsymbol{q}}_M^{(2)}\right) \hat{\boldsymbol{q}}_S^{(2)} U_\mathbf{F}\left(\hat{\boldsymbol{q}}_S^{(2)}, \hat{\boldsymbol{q}}_M^{(2)}\right)\right) \hat{P}_{-1}\left(\hat{q}_{Mz}^{(2)}\right) = U_{\mathbf{F}''}^\dagger\left(\hat{\boldsymbol{q}}_{S,1}^{(2)}\right) \hat{\boldsymbol{q}}_{S,-1}^{(2)} U_{\mathbf{F}''}\left(\hat{\boldsymbol{q}}_{S,1}^{(2)}\right).\end{aligned}\right\} \tag{4.4}$$

Where the unitaries are defined as $U_{\mathbf{F}'}\left(\hat{\boldsymbol{q}}_{S,1}^{(2)}\right) \stackrel{\text{def}}{=} \left((\alpha+\beta)\hat{1}_{S,1} + (\gamma+\delta)\hat{q}_{(S,1)z}^{(2)}\right)$ and $U_{\mathbf{F}''}\left(\hat{\boldsymbol{q}}_{S,-1}^{(2)}\right) \stackrel{\text{def}}{=} \left((\alpha-\beta)\hat{1}_{S,-1} + (\gamma-\delta)\hat{q}_{(S,1)z}^{(2)}\right)$. As can be readily deduced from (4.4), $\mathfrak{Q}_{S,1}$ and $\mathfrak{Q}_{S,-1}$ evolve independently, as the instances perform the unconditional single-qubit rotations **F'** and **F''**, respectively. It is similarly possible to express $U_\mathbf{F}$ in terms of the relative descriptors of $\mathfrak{Q}_{M,1}$ and $\mathfrak{Q}_{M,-1}$ from which it follows, in like manner, that those relative descriptors evolve autonomously under gate **F** too.

Moreover, if the network enacts a single-qubit rotation of $\mathfrak{Q}_M$ or $\mathfrak{Q}_S$ or both (separately), then the relative states evolve independently as well, as such gates do not effect interactions between the qubits and so cannot unentangle them. Thus, the most general gate that ensures the relative states of $\mathfrak{M}$ evolve autonomously between $t=2$ and $t=3$ are products of a gate of the form **F** followed by a single qubit rotation of either or both qubits. This also makes evident the inherently approximate nature of the relative-state construction: in practice, descriptors will only ever be approximately aligned and the system's dynamics will only approximately preserve this alignment, so a foliation into Everett universes is never exact.

As we mentioned in §2, the relative-state instances of both $\mathfrak{Q}_M$ and $\mathfrak{Q}_S$ can store locally inaccessible information. Consider, for instance, the case in which the gate $R_z(\theta)$ acts on $\mathfrak{Q}_M$ between $t = 2$ and $t = 3$, after which the relative descriptors of $\mathfrak{Q}_{M,1}$ and $\mathfrak{Q}_{M,-1}$ depend on the angle $\theta$:

$$\left.\begin{aligned}\hat{\boldsymbol{q}}_{M,1}^{(3)} &= \left(\cos\theta\,\hat{q}_{Mx} - \sin\theta\,\hat{q}_{My},\ \cos\theta\,\hat{q}_{My} + \sin\theta\,\hat{q}_{Mx},\ \hat{q}_{Mz}\right)\hat{P}_1(\hat{q}_{Sx}), \\ \hat{\boldsymbol{q}}_{M,-1}^{(3)} &= \left(\cos\theta\,\hat{q}_{Mx} - \sin\theta\,\hat{q}_{My}, -\cos\theta\,\hat{q}_{My} - \sin\theta\,\hat{q}_{Mx}, -\hat{q}_{Mz}\right)\hat{P}_{-1}(\hat{q}_{Sx}).\end{aligned}\right\} \quad (4.5)$$

The angle $\theta$ does not appear in any of the expectation values of the relative descriptors since $\left\langle\hat{\boldsymbol{q}}_{M,1}^{(3)}\right\rangle_{M,1} = (0,0,1)$ and $\left\langle\hat{\boldsymbol{q}}_{M,-1}^{(3)}\right\rangle_{M,-1} = (0,0,-1)$, implying that both $\mathfrak{Q}_{M,1}$ and $\mathfrak{Q}_{M,-1}$ must contain information that is not accessible through measurements of either qubit instance alone. However, this information could be retrieved through an interference experiment performed on both branches simultaneously, so the qubit instances store locally inaccessible information. This further exemplifies the fact that a relative state represents a fully quantum system.

## 5. Quasi-classical universes

An Everett universe is a quantum system. However, when *some* of the relative descriptors that represent an Everett universe are sharp (with respect to the relative Heisenberg state) and obey quasi-classical equations of motion, then that subset of descriptors represents a *quasi-classical system*. An important example of this is when the relative states of a quantum computational network have sharp z-observables that perform a classical computation, as those observables then behave identically to the computational variables of a classical network (Deutsch (2002)). It is in this sense that Everett 'universes' can resemble a collection of quasi-classical systems (though the spatial extent of these universes never grows faster than the speed of light, and though the relative descriptors that represent them do not jointly specify all the degrees of freedom of the Everettian 'universes').

To illustrate a foliation into quasi-classical systems, consider, for instance, a quantum computational network $\mathfrak{C}$ consisting of $2n$ interacting qubits $\mathfrak{Q}_1, \mathfrak{Q}_2, \ldots, \mathfrak{Q}_{2n}$. We shall treat $\mathfrak{C}$ as two sub-networks $\mathcal{M}$ and $\mathcal{S}$ that consist of the qubits $\mathcal{Q}_1, \mathcal{Q}_2, \ldots, \mathcal{Q}_n$ and $\mathcal{Q}_{n+1}, \mathcal{Q}_{n+2}, \ldots, \mathcal{Q}_{2n}$, respectively. We define

$$\hat{b}_M^{(t)} \stackrel{\text{def}}{=} 2^{n-1} \hat{P}_{-1}\left(\hat{q}_{nz}^{(t)}\right) + \cdots + 2\hat{P}_{-1}\left(\hat{q}_{2z}^{(t)}\right) + \hat{P}_{-1}\left(\hat{q}_{1z}^{(t)}\right). \tag{5.1}$$

Since the projectors in (5.1) each have eigenvalues 0 and 1, the eigenvalues of $\hat{b}_M^{(t)}$ are binary numbers of length $n$ – i.e. elements of $Z_{2^n}$, corresponding to memory states of a classical computer. All these memory states, at time $t$, are simultaneously represented by $\hat{b}_M^{(t)}$, so when the network performs a classical computation (see below), this descriptor represents the evolution of an *ensemble* of classical computers.

Note that $\hat{b}_M^{(t)}$ does not specify the full state of $\mathcal{M}$ at time $t$ but only the state of the $z$-observables of $\left\{\hat{q}_1^{(t)}, \ldots, \hat{q}_n^{(t)}\right\}$. And since those remaining descriptors of $\mathcal{M}$ could, for instance, store locally inaccessible information not present in $\hat{b}_M^{(t)}$, the fully multiversal system $\mathcal{M}$ contains more structure than the ensemble of classical computers (or classical 'universes') represented by $\hat{b}_M^{(t)}$.

We define a similar descriptor for the sub-network $\mathcal{S}$:

$$\hat{b}_S^{(t)} \stackrel{\text{def}}{=} 2^{n-1} \hat{P}_{-1}\left(\hat{q}_{(2n)z}^{(t)}\right) + \cdots + 2\hat{P}_{-1}\left(\hat{q}_{(n+2)z}^{(t)}\right) + \hat{P}_{-1}\left(\hat{q}_{(n+1)z}^{(t)}\right). \tag{5.2}$$

So $\hat{b}_S^{(t)}$ has a spectrum of eigenvalues that is equivalent to $Z_{2^n}$, just as $\hat{b}_M^{(t)}$ does. We shall consider the case in which $\hat{b}_M^{(t)}$ can be foliated into an ensemble of classical computers. To that end, let us assume that at the initial time $t$, sub-network $\mathcal{M}$ is entangled with $\mathcal{S}$. In particular, we assume that the product $\hat{b}_M^{(t)} \hat{b}_S^{(t)}$ is sharp while separately $\hat{b}_S^{(t)}$ and $\hat{b}_M^{(t)}$ are non-sharp, as then, following §3, the descriptor $\hat{b}_M^{(t)}$ decomposes into relative descriptors, each relative to an eigenvalue $j$ of $\hat{b}_S^{(t)}$, i.e.

$$\hat{b}_{M,j}^{(t)} \stackrel{\text{def}}{=} \hat{b}_M^{(t)} \hat{P}_j\left(\hat{b}_S^{(t)}\right) \qquad (j \in Z_{2^n}). \tag{5.3}$$

Here, the projector $\hat{P}_j\left(\hat{b}_S^{(t)}\right)$ projects onto the eigenstate corresponding to the eigenvalue $j$ of $\hat{b}_S^{(t)}$, and the relative Heisenberg-states, associated to each relative descriptor, are normalised products of the projectors $\left\{\hat{P}_j\left(\hat{b}_S^{(t)}\right)\right\}$ and the Heisenberg state $|\Psi\rangle$.

$\mathcal{M}$ is said to perform a reversible classical computation during the $(t+1)$'th computational step if $\hat{b}_M^{(t+1)} = f_t\left(\hat{b}_M^{(t)}\right)$ for some function $f_t$ that maps the spectrum of binary number eigenvalues of $\hat{b}_M^{(t)}$ to itself.[6] And if, between times $t$ and $t+1$, $\mathcal{M}$ effects a classical computation without interacting with $\mathcal{S}$, the relative descriptors (5.3) evolve autonomously:

$$\hat{b}_{M,j}^{(t+1)} = f_t\left(\hat{b}_{M,j}^{(t)}\right) \qquad (j \in Z_{2^n}). \tag{5.4}$$

As can be ascertained from (5.4), the state of $\hat{b}_{M,j}^{(t+1)}$ depends exclusively on the function $f_t$ and the initials descriptor $\hat{b}_{M,j}^{(t)}$, and thus, if $\left\langle \hat{P}_j\left(\hat{b}_S^{(t)}\right)\right\rangle$ is non-zero, then during the $(t+1)$'th computational step the information contained in $\hat{b}_{M,j}^{(t)}$ is processed by that relative descriptor alone. Moreover, each relative descriptor $\hat{b}_{M,j}^{(t)}$ has a sharp value with respect to its relative Heisenberg state and has a unique history defined by the function $f_t$. Which is to say that the relative descriptors $\hat{b}_{M,j}^{(t)}$ describe simultaneously existing classical systems.

## 6. Conclusions

We have formulated the relative-state construction in the Heisenberg picture. The relative-state foliation is necessarily local physically, and physically, the Everett multiplicity can only spread at the speed of light or less, and the spatially finite Everett universes are fully quantum.

## Acknowledgements

We are grateful for perceptive criticisms of earlier drafts of this paper by Chiara Marletto, Vlatko Vedral, Eric Marcus, David Felce, and Lev Vaidman. This work was supported in part by the Prins Bernhard Cultuurfonds.

## References

---

[6] The function $f_t$ can invariably be implemented by a network consisting exclusively of **ccnot** (Toffoli) gates, as this gate is universal for reversible classical computation (Deutsch (2002)).

Byrne, P. (2010) *The Many Worlds of Hugh Everett III: Multiple Universes, Mutual Assured Destruction, and the Meltdown of a Nuclear Family.* Oxford University Press.

Deutsch, D. (2002) *The Structure of the Multiverse*. Proc. R. Soc. Lond. **A458**, 2911-2923. (doi:10.1098/rspa.2002.1015)

Deutsch, D. (2011) *Vindication of Quantum Locality*. Proc. R. Soc. Lond **A468**, 531-544. (doi:10.1098/rspa.2011.0420)

Deutsch, D. & Hayden, P. (2000*) Information Flow in Entangled Quantum Systems.* Proc. R. Soc. Lond. **A456**, 1759–1774. (doi:10.1098/rspa.2000.0585)

DeWitt, B.S. & Graham, N. (1973) *The Many Worlds Interepretation of Quantum Mechanics.* Princeton University Press, Princeton.

Dirac, P.A.M. (1965) *Quantum Electrodynamics without Dead Wood.* Phys. Rev. **139**, B684-B690. (doi:10.1103/PhysRev.139.B684)

Everett (1973) *Theory of the Universal Wave Function* in DeWitt & Graham *loc. cit.*

Gottesman, D. (1999) *Group22: Proceedings of the XXII International Colloquium on Group Theoretical Methods in Physics*, S. P. Corney, R. Delbourgo, and P. D. Jarvis, (eds.) 32-43 International Press, Cambridge, MA.

Hewitt-Horsman, C. & Vedral, V. (2007a) *Developing the Deutsch–Hayden approach to quantum mechanics.* New J. Phys. **9**, 135. (doi:10.1088/1367-2630/9/5/135)

Hewitt-Horsman, C. & Vedral, V. (2007b) *Entanglement without nonlocality.* Phys. Rev. **A76** 062319. (doi:10.1103/PhysRevA.76.062319)

Vaidman, L., *Many-Worlds Interpretation of Quantum Mechanics*, Stanford Encyclopedia of Philosophy (Fall 2018 Edition), Edward N. Zalta (ed.).

Wallace, D. (2011) *The Everett Interpretation.* Oxford University Press.

Wallace, D. (2012a) *Decoherence and its role in the modern measurement problem.* Phil. Trans. R. Soc. **A370**, 4576-4593. (doi:10.1098/rsta.2011.0490)

Wallace, D. (2012b) *The Emergent Multiverse.* Oxford University Press.

Zurek, W.H. (1981) *Pointer basis of quantum apparatus: Into what mixture does the wave packet collapse?* Phys. Rev. **D24**, 1516-1525.

Zurek, W.H. (1982) *Environment-induced superselection rules.* Phys. Rev. **D26**, 1862-1880. (doi:10.1103/PhysRevD.26.1862)